\documentclass[submission,copyright,creativecommons,fleqn]{eptcs}

\providecommand{\event}{EXPRESS/SOS 2017} 

\usepackage{breakurl}             
\usepackage{hyperref}
\hypersetup{bookmarks=true,bookmarksnumbered=true,bookmarksopen}
\usepackage{amsmath}
\usepackage{amsthm}
\usepackage[usenames]{color}
\usepackage{amssymb}
\usepackage{procalg}
\usepackage{tikz}
\usetikzlibrary{arrows,automata}
\usepackage{amscd}
\usepackage{txfonts}
\usepackage{amsfonts,semantic,colortbl,mathrsfs,stmaryrd,mathtools}
\usepackage{epsfig,graphicx,subfigure}
\usepackage{enumerate,enumitem}
\usepackage{multirow,multicol}
\usepackage{tabularx}
\usepackage{gastex}
\usepackage{caption}
\usepackage{color}
\makeatletter
\@ifundefined{ifAP}{%

\newcommand{\todo}[1]{\underline{\sf todo}: {\color{red}{#1}}}
}{
\ifAP
\else
\fi
\newcommand{\todo}[1]{}
}
\makeatother

\newenvironment{tarray}[2][c]{
  \settowidth{\dimen1}{${} = {}$}%
  \setlength{\arraycolsep}{0.125\dimen1}%
  \hspace{-0.125\dimen1}\array[#1]{#2}\relax
}
{
   \endarray\hspace{-0.125\dimen1}
}
\def\tbb{\begin{tarray}{llll}}
\def\tbbt{\begin{tarray}[t]{llll}}
\def\tee{\end{tarray}}

\renewcommand{\bullet}{\ensuremath{;}}

\newcommand{\T}{\ensuremath{\mathcal{T}} }
\newcommand{\A}{\ensuremath{\mathcal{A}}}
\newcommand{\Atau}{\ensuremath{\mathcal{A}_{\tau}}}
\newcommand{\Sta}{\ensuremath{\mathcal{S}}}

\newcommand{\R}{\mathcal{\mathop{R}}}
\newcommand{\M}{\mathcal{M}}
\newcommand{\F}{\mathcal{F}}
\newcommand{\D}{\mathcal{D}}
\newcommand{\N}{\mathcal{N}}
\newcommand{\I}{\mathcal{I}}
\newcommand{\C}{\mathcal{C}}
\newcommand{\V}{\mathcal{V}}
\newcommand{\ddef}{\overset{\textrm{def}}{=}}
\newcommand{\Dbox}{\mathcal{D}_{\Box}}
\newcommand{\tphd}[1]{\check{#1}}
\newcommand{\tphdL}[1]{{#1 \!}^{\scriptscriptstyle <}}
\newcommand{\tphdR}[1]{\prescript{\scriptscriptstyle >}{}{\! #1}}
\newcommand{\Conf}[1][]{\mathcal{C}_{#1}}
\newcommand{\length}[1]{\ensuremath{\mathit{length}(#1)}}
\newcommand{\stack}[1]{\ensuremath{\mathit{stack}(#1)}}
\newcommand{\get}[2]{\ensuremath{\mathit{get}(#1,#2)}}
\newcommand{\suffset}[2]{\ensuremath{\mathit{suffset}(#1,#2)}}
\newcommand{\ar}{\ensuremath{\mathit{ar}}}
\newcommand{\bbisimd}{\ensuremath{\mathrel{\bbisim^{\Delta}}}}
\newcommand{\rbbisimd}{\ensuremath{\mathrel{\rbbisim^{\Delta}}}}
\newcommand{\nil}{\ensuremath{\mathalpha{\mathbf{0}}}}
\newcommand{\one}{\ensuremath{\mathalpha{\mathbf{1}}}}

\newcommand{\encode}[1]{\lceil #1\rceil}

\newcommand{\mer}{\ensuremath{\mathit{merge}}}

\newcommand{\nesting}{\ensuremath{\mathalpha{\sharp}}}
\newcommand{\pushdown}{\ensuremath{\mathalpha{\$}}}
\newcommand{\backforth}{\ensuremath{\mathalpha{\leftrightarrows}}}

\newcommand{\TCP}{\ensuremath{\textrm{TCP}}}
\newcommand{\TCPN}{\ensuremath{\textrm{TCP}^{\nesting}}}
\newcommand{\TCPP}{\ensuremath{\textrm{TCP}^{\pushdown}}}
\newcommand{\TCPB}{\ensuremath{\textrm{TCP}^{\backforth}}}

\newcommand{\TCPS}{\ensuremath{\textrm{TCP}^{\bullet}}}

\newcommand{\SA}{\ensuremath{\textrm{TSP}}}
\newcommand{\SAS}{\ensuremath{\textrm{TSP}^{\bullet}}}

\newcommand{\PDA}{\ensuremath{\textrm{PDA}}}

\newcommand{\pushd}[2]{\ensuremath{\mathalpha{{#1}^{\pushdown}{#2}}}}
\newcommand{\backf}[2]{\ensuremath{\mathalpha{{#1}^{\backforth}{#2}}}}
\newcommand{\nest}[2]{\ensuremath{\mathalpha{{#1}^{\nesting}{#2}}}}
\newcommand{\delete}[1]{}

\newcommand{\FY}[1]{{#1}} 
\newcommand{\conf}[1]{#1}
\newcommand{\arx}[1]{}

\newtheorem{definition}{Definition}
\newtheorem{lemma}{Lemma}

\newtheorem{theorem}{Theorem}

\title{Sequential Composition in the Presence of Intermediate Termination (Extended Abstract)}
\author{Jos Baeten
\institute{CWI\\ Amsterdam, the Netherlands}
\institute{University of Amsterdam,\\ Amsterdam, the Netherlands}
\email{Jos.Baeten@cwi.nl}
\and
Bas Luttik
\institute{Eindhoven University of Technology\\ Eindhoven, the Netherlands}
\email{s.p.luttik@tue.nl}
\and
Fei Yang
\institute{Eindhoven University of Technology\\ Eindhoven, the Netherlands}
\email{\quad f.yang@tue.nl}
}
\def\titlerunning{Sequential Composition in the Presence of Intermediate Termination}
\def\authorrunning{Jos Baeten, Bas Luttik \& Fei Yang}

\begin{document}
\maketitle

\begin{abstract}
The standard operational semantics of the sequential composition operator gives rise to unbounded branching and forgetfulness when transparent process expressions are put in sequence. Due to transparency, the correspondence between context-free and pushdown processes fails modulo bisimilarity, and it is not clear how to specify an always terminating half counter. We propose a revised operational semantics for the sequential composition operator in the context of intermediate termination.  With the revised operational semantics, we eliminate transparency\FY{, allowing us to} establish a \FY{close} correspondence between context-free processes and pushdown processes. Moreover, we prove the reactive Turing powerfulness of TCP with iteration and nesting with the revised operational semantics for sequential composition.
\end{abstract}

\section{Introduction}\label{sec:introduction}
\delete{\todo{\begin{enumerate}
\item We need to illustrate the origin of the old version of sequential composition operator.
\item We shall illustrate the role of termination.
\item We shall illustrate the problems of the old version: transparency.
\item We shall introduce the problems of pushdown automata and context free process and reactive Turing powerful process calculus and our results.
\end{enumerate}}
The integration of concurrency theory and the classical theory of formal languages has been extensively studied in recent years~\cite{BCLT2009}. A lot of notions in the classical theory have their counterparts in concurrency theory~\cite{vanTilburg2011}. However, we still cannot conclude a complete correspondence for all the notions. As far as we are concerned, a major obstacle is the phenomenon of transparency due to intermediate termination in the context of sequential composition.
}
Sequential composition is a standard operator in many process calculi. The functionality of the sequential composition operator is to concatenate the behaviours of two systems. It has been widely used in many process calculi with the notation ``$\cdot$''. We illustrate its operational semantics by a process $P\cdot Q$ in \TCP~\cite{BBR2010}. If the process $P$ has a transition $P\step{a}P'$ for some action label $a$, then the composition $P\cdot Q$ has the transition $P\cdot Q\step{a}P'\cdot Q$.
Termination is an important behaviour for models of computation~\cite{BBR2010}. \FY{A semantic distinction between successful and unsuccessful termination in concurrency theory (CT) is especially important for a smooth incorporation of the classical theory of automata and formal languages (AFT): the distinction is used to express whether a state in an automaton is accepting or not. Automata may even have states that are accepting and may still perform transitions; this phenomenon we call \emph{intermediate termination}. From a concurrency-theoretic point of view, such behaviour is perhaps somewhat unnatural. To be able to express it nevertheless, we let an alternative composition inherit} the option to terminate from just \FY{one} of its components. \FY{The expression} $a.(b+1)$ \FY{then denotes} the process that does an $a$-transition and \FY{subsequently} enters a state that is successfully terminated but can also do a $b$-transition.

 \FY{To specify the operational semantics of sequential composition in a setting with a explicit successful termination, usually the following three rules are added: the} first one states that \FY{the sequential composition} $P\cdot Q$ terminates if both $P$ and $Q$ terminate; \FY{the second one states that if $P$ admits a transition $P\step{a}P'$, then $P\cdot Q$ admits a transition $P\cdot Q\step{a}P'\cdot Q$;} and the \FY{third} one states that if $P$ terminates, and there is a transition $Q\step{a}Q'$, then we have the transition $P\cdot Q\step{a}Q'$.

In this paper, we discuss a complication \FY{stemming from these} operational semantics of the sequential composition operator. \FY{The complication is that a process expression $P$ with the option to terminate is \emph{transparent} in a sequential context $P\cdot Q$: if $P$ may still perform observable behaviour other than termination, then this may be skipped by doing a transition from $Q$.} \FY{There are} two disadvantages of transparency in our attempts to achieve a smooth integration of process theory and the classical theory of automata and formal languages \cite{BCLT2009}:

The relationship between context-free processes \FY{(i.e., processes that can be specified with a guarded recursive specification over a language with action
constants, constants for deadlock ($\nil$) and successful termination ($\one$), and binary operations for sequential and alternative composition)} and pushdown \FY{automata} has been extensively discussed in the literature~\cite{BCvT2008}. It has been shown that \FY{every context-free process is equivalent to the behaviour of some pushdown automaton (i.e., a pushdown process)} modulo contra simulation\FY{, but not }modulo rooted branching bisimulation. By stacking unboundedly many transparent terms with sequential composition, we would get an unboundedly branching transition system. It was shown that unboundedly branching behaviour cannot be specified by any pushdown process modulo rooted branching bisimulation~\cite{BCvT2008}. In order to improve the result to a finer notion of behaviour equivalence, we need to eliminate the problem of unbounded branching.

Transparency also complicates matters if one wants to specify some form of memory (e.g., a counter, a stack, or a tape) that always has the option to terminate, but at the same time does not lose data. If the standard process algebraic specifications of such memory processes are generalised to a setting with intermediate termination, then either they are not always terminating, or they are `forgetful' and may non-deterministically lose data. This is a concern when one tries to specify the behaviour of a pushdown automaton or a Reactive Turing machine in a process calculus \cite{BLT2013, LY15, LY16}. The process calculus \TCP{} with iteration and nesting is Turing complete~\cite{Bergstra1994,bergstra2001non}. Moreover, it follows from the result in~\cite{bergstra2001non} that it is reactively Turing powerful if intermediate termination is not considered. However, it is not clear to us how to reconstruct the proof of reactive Turing powerfulness if termination is considered. Due to the forgetfulness on the stacking of transparent process expressions, it is not clear to us how to define a counter that is always terminating, which is crucial \FY{for establishing} the reactive Turing powerfulness.

In order to avoid the (in some cases) undesirable feature of unbounded branching and forgetfulness, we propose a revised operational semantics for the sequential composition operator. The modification consists of disallowing a transition from the second component of a sequential composition if the first component is able to perform a transition. Thus, we avoid the problems mentioned above with the revised operator. We shall prove that every context-free process is bisimilar to a pushdown process, and that \TCP{} with iteration and nesting is reactively Turing powerful modulo divergence-preserving branching bisimilarity (without resorting to recursion) in the revised semantics.

\FY{The research presented in this article is part of an attempt to achieve a smoother integration of the classical theory of automata and formal languages (AFT) within concurrency theory (CT). The idea is to recognise that a finite automaton is just a special type of labelled transition system, that more complicate automata (pushdown automata, Turing machines) naturally generate transition systems, and that there is a natural correspondence between regular expressions and grammars on the one hand and certain process calculi on the other hand. In~\cite{BCLT2009,BLT11a,BLT12} we have studied the various notions of automata from AFT modulo branching bisimilarity. In~\cite{BLMvT16} we have explored the correspondence between finite automata and regular expressions extended with parallel composition modulo strong bisimilarity. In~\cite{BLT2013} we have proposed reactive Turing machines as an extension of Turing machines with concurrency-style interaction.}

The paper is structured as follows. We first introduce TCP with the standard version of sequential composition in Section~\ref{sec:preliminaries}. Next, we discuss the complications caused by transparency in Section~\ref{sec:transparency}. Then, in Section~\ref{sec:sequential}, we propose the revised operational semantics of the sequential composition operator, and show that rooted divergence-preserving branching bisimulation is a congruence. In Section~\ref{sec:cfg}, we revisit the relationship between context-free processes and pushdown \FY{automata}, and show that every context-free process is bisimilar to a pushdown process in our revised semantics. In Section~\ref{sec:Termination}, we prove that \TCP{} with iteration and nesting is reactively Turing powerful in the revised semantics. In Section~\ref{sec:conclusion}, we draw some conclusions and propose some future work.
\conf{\FY{The full version of this extended abstract, including proofs of the results, is available as~\cite{LY17a}.}}

\section{Preliminaries}\label{sec:preliminaries}
We start with introducing \FY{the} notion of labelled transition system, which is used as the standard mathematical representation of behaviour. We consider transition systems with a subset of states marked \FY{as terminating} states. We let $\A$ be a set of \emph{action symbols}, and we extend $\A$ with a special symbol $\tau\notin \A$, which intuitively denotes unobservable internal activity of the system. We shall abbreviate $\A \cup\{\tau\}$ by $\Atau$.
\begin{definition}
~\label{def:lts}
An \emph{$\Atau$-labelled transition system} is a tuple $(\Sta,\step{},\uparrow,\downarrow)$, where
\begin{enumerate}
    \item $\Sta$ is a set of \emph{states},
    \item ${\step{}}\subseteq{\Sta\times\Atau\times\Sta}$ is an $\Atau$-labelled \emph{transition relation},
    \item ${\uparrow}\in\Sta$ is the initial state, and
    \item ${\downarrow}\subseteq\Sta$ is a set of terminating states.
\end{enumerate}
\end{definition}
\delete{
\begin{definition}
A \emph{signature} $\Sigma$ consists of:
\begin{enumerate}
\item an infinite set of variables $x,y,z,\ldots$; and
\item a set $\F$ of function symbols $f,g,h,\ldots$, where each symbol $f$ has an arity $\ar(f)$.
\end{enumerate}
A function symbol of arity zero is called a \emph{constant}.
\end{definition}

\begin{definition}
Let $\Sigma$ be a signature. The collection $\mathbb{T}(\Sigma)$ of \emph{(open) terms} $p,q,r,\ldots$ over $\Sigma$ is defined as the least set satisfying:
\begin{enumerate}
\item each variable is in $\mathbb{T}(\Sigma)$;
\item if $t_1,\ldots,t_{\ar(f)}\in\mathbb{T}(\Sigma)$, then $f(t_1,\ldots,t_{\ar(f)})\in\mathbb{T}(\Sigma)$.
\end{enumerate}
A term is \emph{closed} if it does not contain any free variables. The set of closed terms is denoted by $\mathbb{C}(\Sigma)$.
\end{definition}
}

Next, we shall introduce the process calculus \emph{Theory of Sequential Processes} (\SA) that allows us to describe transition systems. 

\FY{Let} $\N$ be a countably infinite set of names. The set of process expressions $\mathcal{P}$ is generated by the following grammar $(a\in\Atau,\,N\in\N)$:
\begin{equation*}
P:=\nil\mid \one\mid a.P\mid P\cdot P\mid P+P\mid N
\enskip.
\end{equation*}

We briefly comment on the operators in this syntax. The constant $\nil$ denotes \emph{deadlock}, the unsuccessfully terminated process. The constant $\one$ denotes \emph{termination}, the successfully terminated process. For each action $a\in\Atau$ there is a unary operator $a.$ denoting action prefix; the process denoted by $a.P$ can do an $a$-labelled transition to the process $P$. The binary operator $+$ denotes alternative composition or choice. The binary operator $\cdot$ represents the sequential composition of two processes.
\delete{
Let $P$ be an arbitrary process expression; and we use an abbreviation inductively defined by:
\begin{enumerate}
\item $P^{0}=\one$; and
\item $P^{n+1}=P\cdot P^{n}$ for all $n\in\mathbb{N}$.
\end{enumerate}}

Let $P$ be an arbitrary process expression; and we use an abbreviation inductively defined by: $P^{0}=\one$; and $P^{n+1}=P\cdot P^{n}$ for all $n\in\mathbb{N}$.

A recursive specification $E$ is a set of equations
$E = \{N\ddef P | N\in \N, P\in\mathcal{P}\}$,
satisfying:
\begin{enumerate}
\item for every $N\in\N$ it includes at most one equation with $N$ as left-hand side,
which is referred to as the \emph{defining equation} for $N$; and
\item if some name $N'$ occurs in the right-hand side $P'$ of some equation $N' = P'$ in
$E$, then $E$ must include a defining equation for $N'$.
\end{enumerate}

\FY{
An occurrence of a name $N$ in a process expression is \emph{guarded} if the occurrence is within the scope of an action prefix $a.$ for some $a\in A$ ($\tau$ cannot be a guard). A recursive specification $E$ is guarded if all occurrences of names in right-hand sides of equations in $E$ are guarded.}

We use structural operational semantics to associate a transition relation with process expressions defined in \SA.  A term is \emph{closed} if it does not contain any free variables. Structural operational semantics induces a transition relation on closed terms.   We let $\step{}$ be the $\Atau$-labelled transition relation induced on the set of process expressions by operational rules in Figure~\ref{fig:semantics-tsp}. Note that we presuppose a recursive specification $E$, and we omit the symmetrical rules \FY{for $+$}.

\begin{figure}[h]
\begin{center}
\fbox{
\begin{minipage}[t]{.7\textwidth}
\begin{eqnarray*}
&\inference{}{\one\downarrow}\quad\inference{\,}{a.P\step{a}P}\quad
\inference{P_1\step{a}P_1'}{P_1+ P_2\step{a}P_1'}\quad
\inference{P_1\downarrow}{P_1+P_2\downarrow}\\
& \inference{P_1\downarrow\quad  P_2\downarrow}{P_1\cdot P_2\downarrow}\quad
\inference{P_1\step{a}P_1'}{P_1\cdot P_2\step{a}P_1'\cdot P_2}\quad \inference{P_1\downarrow\quad P_2\step{a}P_2'}{P_1\cdot P_2\step{a}P_2'}\\
&\inference{P\step{a}P'\quad (N\ddef P)\in E}{N\step{a}P'}\quad\inference{P\downarrow\quad (N\ddef P)\in E}{N\downarrow}
\end{eqnarray*}
\end{minipage}
}\end{center}
\caption{The operational semantics of TSP}\label{fig:semantics-tsp}
\end{figure}
Here we use $P\step{a}P'$ to denote an $a$-labelled transition $(P,a,P')\in{\step{}}$. We say a process expression $P'$ is \emph{reachable} from $P$ is there exist process expressions $P_0,\ldots,P_n$ and labels $a_1,\ldots,a_n$ such that $P=P_0\step{a_1}\FY{\cdots}\step{a_n}P_n=P'$.

Given a \SA{} process expression $P$, the transition system $\T(P)=(\Sta_P,\step{}_P,\uparrow_P,\downarrow_P)$ associated with $P$ is defined as follows:
\begin{enumerate}
\item the set of states $\Sta_P$ consists of all process expressions reachable from $P$;
\item the \FY{transition relation} $\step{}_P$ is the restriction to $\Sta_P$ of the transition relation defined on all process expressions by the structural operational semantics, i.e., ${\step{}_P}={\step{}}\cap{(\Sta_P\times\Atau\times\Sta_P)}$;
\item ${\uparrow_P}=P$; and
\item the set of final states $\downarrow_P$ consists of all process expressions $Q\in\Sta_P$ such that $Q\downarrow$, i.e., ${\downarrow_P}={\downarrow\cap\Sta_P}$.
\end{enumerate}

We also use \FY{(a restricted variant of)} the process calculus \TCP{} in later sections. It is obtained by adding a parallel composition operator \FY{to} \SA.
Let $\C$ be a set of \emph{channels} and $\Dbox$ be a set of \emph{data symbols}. For every subset $\C'\subseteq\C$, we \FY{propose} a special set of actions $\I_{\C'}\subseteq \Atau$ \FY{defined} by: $\I_{\C'}=\{c?d,c!d\mid d\in\Dbox,c\in \C'\}.$

The actions $c?d$ and $c!d$ denote the events that a datum $d$ is received or sent along channel $c$\FY{, respectively}. \FY{We include binary} parallel composition operators \FY{$[\_ \| \_]_{\C'}$ ($\C′\subseteq \C$)}.
\FY{Communication along the channels in $\C'$ is enforced} and communication results in $\tau$.

The operational semantics of the parallel composition operators is \FY{presented} in Figure~\ref{fig:parallel-composition} (We omit the symmetrical rules).
\FY{%
\begin{figure}[h]
\begin{center}
\fbox{
\begin{minipage}[t]{0.9\textwidth}
\begin{eqnarray*}
&\inference{P_1\step{a}P_1'\quad a\notin \I_{\C'}}{[P_1\parallel P_2]_{\C'}\step{a}[P_1'\parallel P_2]_{\C'}}\quad\inference{P_1\downarrow\quad P_2\downarrow}{[P_1\parallel P_2]_{\C'}\downarrow}
\quad\inference{P_1\step{c?d}P_1'\quad P_2\step{c!d}P_2'\quad c\in\C'}{[P_1\parallel P_2]_{\C'}\step{\tau}[P_1'\parallel P_2']_{\C'}}
\end{eqnarray*}
\end{minipage}
}\end{center}
\caption{The operational semantics of parallel composition}\label{fig:parallel-composition}
\end{figure}}

The notion of behavioural equivalence has been used extensively in the theory of process calculi.
We first introduce the notion of strong \FY{bisimilarity}~\cite{M1989,P1981}, which does not distinguish $\tau$-transitions from other labelled transitions.

\begin{definition}~\label{def:bisim}
A binary symmetric relation $\R$ on a transition system $(\Sta,\step{},\uparrow,\downarrow)$ is a \emph{strong bisimulation} if, for all states $s,t\in\Sta$, $s\R t$ implies
\begin{enumerate}
 \item if $s\step{a}s'$, then there exist $t'\in\Sta$, such that $t\step{a}t'$, and $s'\R t'$;
    \item if $s\downarrow$, then $t\downarrow$.
\end{enumerate}
The states $s$ and $t$ are \emph{strongly bisimilar} (notation: $s\bisim t$) if there exists a strong bisimulation $\R$ s.t. $s\R t$.
\end{definition}

The notion of strong bisimilarity does not take into account the intuition associated with $\tau$ that it stands for unobservable internal activity. We proceed to introduce the notion of (divergence-preserving) branching bisimilarity, which does treat $\tau$-transitions as unobservable. Divergence-preserving branching bisimilarity is the finest behavioural equivalence in van Glabbeek's linear time - branching time spectrum~\cite{Glabbeek1993}, \FY{and, moreover, the coarsest behavioural equivalence compatible with parallel composition that preserves validity of formulas from the well-known modal logic CTL minus the next-time modality $X$~\cite{vGLT2009b}.}
Let $\step{}$ be an $\Atau$-labelled transition relation on a set $\Sta$, and let $a\in\Atau$; we write $s\step{(a)}t$ for the \FY{formula} ``$s\step{a}t\vee (a=\tau\wedge s=t)$''. Furthermore, we denote the transitive closure of $\step{\tau}$ by $\step{}^{+}$ and the reflexive-transitive closure of $\step{\tau}$ by $\step{}^{*}$.

\begin{definition}
~\label{def:bbisim}
Let $T=(\Sta,\step{},\uparrow,\downarrow)$ be a transition system. A branching bisimulation is a symmetric relation $\R\subseteq\Sta\times\Sta$ such that for all states $s,t\in\Sta$, $s\R t$ implies
\begin{enumerate}
    \item if $s\step{a}s'$, then there exist $t',t''\in\Sta$, such that $t\step{}^{*}t''\step{(a)}t'$, $s\R t''$ and $s'\R t'$;
    \item if $s\downarrow$, then there exists $t'\FY{\in\Sta}$ such that $t\step{}^{*} t'$\FY{, $t'\downarrow$ and $s\R t'$}.
\end{enumerate}
The states $s$ and $t$ are \emph{branching bisimilar} (notation: $s\bbisim t$) if there exists a branching bisimulation $\R$ \FY{such that} $s\R t$.

A branching bisimulation $\R$ is \emph{divergence-preserving} if, for all states $s$ and $t$, $s\R t$ implies
\begin{enumerate}
\setcounter{enumi}{2}
    \item if there exists an infinite sequence $(s_{i})_{i\in\mathbb{N}}$ such that $s=s_{0},\,s_{i}\step{\tau}s_{i+1}$ and $s_{i}\R t$ for all $i\in\mathbb{N}$, then there exists a state $t'$ such that $t\step{}^{+}t'$ and $s_{i}\R t'$ for some $i\in\mathbb{N}$.
    \end{enumerate}
The states $s$ and $t$ are \emph{divergence-preserving branching bisimilar} (notation: $s\bbisimd t$) if there exists a divergence-preserving branching bisimulation $\R$ such that $s\R t$.
\end{definition}

\FY{The relation $\bbisimd$ satisfies the conditions of Definition~\ref{def:bbisim}, and is, in fact, the largest divergence-preserving branching bisimulation relation.} \FY{Divergence-preserving branching bisimilarity is an equivalence relation}~\cite{vGLT2009a}.
\delete{
\begin{definition}
An equivalence relation $\R$ on a process calculus $C$ is called a congruence if $s_i \R t_i$ for $i=1, ..., ar( f )$ implies $f (s_1 ,\ldots, s_{ar( f )})\R f(t_1 ,\ldots, t_{ar(f)})$, where $f$ is an operator of $C$, $ar(f)$ is the arity of $f$, and $s_i,t_i$ are processes defined in $C$.
\end{definition}
}

Divergence-preserving branching \FY{bisimilarity} is not a congruence \FY{for \SA{}; it is well-known that it is not compatible with alternative composition.}. A rootedness condition needs to be introduced.

\begin{definition}~\label{def:root}
Let $T=(\Sta,\step{},\uparrow,\downarrow)$ be a transition system. A divergence-preserving branching bismulation relation $\R$ on $T$ satisfies the \emph{rootedness} condition \FY{for} a pair of states $s_1,s_2\in\Sta$, if $s_1\R s_2$ and
\begin{enumerate}
\item if $s_1\step{a}s_1'$, then $s_2\step{a}s_2'$ for some $s_2'$ such that $s_1'\R s_2'$;
\item if $s_1\downarrow$, then $s_2\downarrow$.
\end{enumerate}
$s_1$ and  $s_2$ are \emph{rooted divergence-preserving branching bisimilar} (notation: $s_1\rbbisimd s_2$) if there exists a divergence-preserving branching bisimulation $\R$ \FY{that} satisfies rootedness condition \FY{for} $s_1$ and $s_2$.
\end{definition}

We can extend the above relations ($\bisim,\,\bbisim,\,\bbisimd,$ and $\rbbisimd$) to relations over two transition systems by \FY{defining that they} are bisimilar if their initial states are bisimilar in \FY{their disjoint union}. Namely, for two transition systems $T_1=(\Sta_1,\step{}_1,\uparrow_1,\downarrow_1)$ and $T_2=(\Sta_2,\step{}_2,\uparrow_2,\downarrow_2)$, we make the following pairing on their states. We pair every state $s\in\Sta_1$ with $1$ and every state $s\in\Sta_2$ with $2$. We have $T_i'=(\Sta_i',\step{}_i',\uparrow'_i,\downarrow'_i)$ for $i=1,2$ where $\Sta_i'=\{(s,i)\mid s\in\Sta_i\}$, $\step{}_i'=\{((s,i),a,(t,i))\mid (s,a,t)\in\step{}_i\}$, $\uparrow_i'=(\uparrow_i,i)$, and $\downarrow_i'=\{(s,i)\mid s\in\downarrow_i\}$. We say $T_1\equiv T_2$ if \FY{in} $T=(\Sta_1'\cup\Sta_2',{\step{}_1'}\cup{\step{}_2'},\uparrow_1',{\downarrow_1'}\cup{\downarrow_2'})$ \FY{we have} $\uparrow_1'\equiv\uparrow_2'$.

\section{Transparency}\label{sec:transparency}
Process expressions that have the option to terminate are \emph{transparent} in a sequential context: if $P$ has the
option to terminate and $Q\step{a}Q'$, then $P\cdot Q\step{a}Q'$ even if $P$ can still do transitions. In this section we shall
explain how transparency gives rise to two phenomena that are undesirable in certain circumstances. First, it
facilitates the specification of unboundedly branching behaviour with a guarded recursive specification over
\SA{}.  Second, it gives rise to forgetful stacking of
variables, and as a consequence it is not clear how to specify an always terminating half-counter.

\begin{figure}
\centering
\begin{tikzpicture}[->,>=stealth',shorten >=1pt,auto,node distance=2cm,semithick]
   \tikzstyle{every state}=[rectangle,rounded corners,draw=black, top color=white, bottom color=yellow!50,very thick, inner sep=0.2cm, minimum size=0.7cm, text centered]
    \node[state,initial] (A)                    {$X$};
  \node[state]         (B) [right of=A] {$X\cdot Y$};
  \node[state]         (C) [right of=B] {$X\cdot Y^2$};
  \node[state]         (D) [right of=C] {$X\cdot Y^{n-1}$};
  \node[state]         (E) [right of=D] {$X\cdot Y^{n}$};
  \node[state,accepting]         (F) [below of=  E]       {$Y^{n}$};
  \node[state,accepting]         (G) [left of=F]       {$Y^{n-1}$};
\node[state,accepting]         (H) [left of=G]       {$Y^2$};
\node[state,accepting]         (I) [left of=H]       {$Y$};
\node[state,accepting]         (J) [left of=I]       {$\one$};
\node[] (K)[right of=E]{};
\node[] (L)[right of=F]{};
  \path (A) edge              node {$a$} (B)
            edge              node {$b$} (J)
        (B) edge            node {$a$} (C)
            edge              node {$b$} (I)
        (C) edge[dashed]              node {} (D)
            edge              node {$b$} (H)
        (D) edge              node {$a$} (E)
            edge              node {$b$} (G)
        (E) edge              node {$b$} (F)
            edge[dashed]      node {} (K)
        (F) edge              node[above] {$c$} (G)
            edge[bend left]    node[above] {$c$} (H)
            edge[bend left]    node[above] {$c$} (G)
            edge[bend left]    node[above] {$c$} (I)
            edge[bend left]    node[above] {$c$} (J)
        (G) edge[dashed]                node{} (H)
            edge[bend left]    node[above] {$c$} (H)
            edge[bend left]    node[above] {$c$} (I)
            edge[bend left]    node[above] {$c$} (J)
        (H) edge      node[above] {$c$} (I)
            edge[bend left]    node[above] {$c$} (J)
        (I) edge      node[above] {$c$} (J)
        (L) edge[dashed]      node {} (F);
\end{tikzpicture}

\caption{A transition system with unboundedly branching behaviour}~\label{fig:unbounded}
\end{figure}
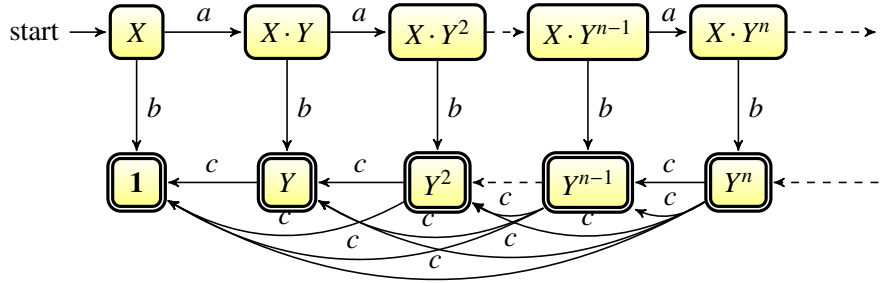

We first discuss process expressions with unbounded branching.
It is well-known from formal language theory that the context-free languages are
exactly the languages accepted by pushdown automata. The process-theoretic formulation of this result is
that every transition system specified by a \SA{} specification is language equivalent to the transition
system associated with a pushdown automaton and, vice versa, every transition system associated with a
pushdown automaton is language equivalent to the transition system associated with some \SA{}
specification. The correspondence fails, however, when language equivalence is replaced by (strong)
bisimilarity. \FY{The currently tightest} result is that for every context-free process there is a pushdown process to simulate it modulo contra simulation~\cite{BCvT2008}\FY{; we conjecture that not every context-free process is simulated by a pushdown process modulo} branching \FY{bisimilarity}. The reason is that context-free processes \FY{may} have \FY{an} unbounded branching degree. Consider the following process:
\begin{equation*}
X=a.X\cdot Y+b.\one\quad Y=c.\one+\one
\enskip.
\end{equation*}
The transition system \FY{associated with $X$} is illustrated in Figure~\ref{fig:unbounded}. Note that every state in the second row is a terminating state. \FY{The} state $Y^n$ has $n$ $c$-labelled transitions to $\one,Y,Y^2,\ldots,Y^{n-1}$, respectively. Therefore, every state in this transition system has finitely many transitions leading to distinct states, \FY{but} there is no upper bound on the number of transitions from each state. \FY{Therefore}, we say that this transition system \FY{has an unbounded branching degree.}

We can prove that the process defined by the \SA{} specification above is not strongly bisimilar to a pushdown process since it \FY{has an unbounded branching degree}, whereas a pushdown process is always boundedly branching. \FY{The} correspondence does hold \FY{modulo} contra simulation~\cite{BCvT2008}, and it is an
open problem as to whether the correspondence holds modulo branching bisimilarity. In Section~\ref{sec:cfg}, we show that with \FY{a revised operational semantics for sequential composition}, we \FY{eliminate} such unbounded branching and \FY{indeed obtain} a correspondence between pushdown processes and context-free processes modulo strong bisimilarity.

\begin{figure}[h]
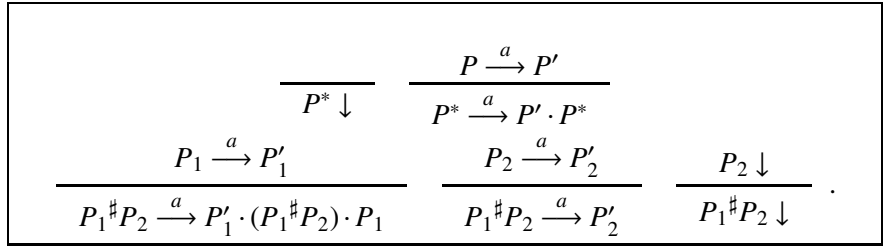

\begin{center}
\fbox{
\begin{minipage}[t]{.7\textwidth}
\begin{eqnarray*}
&\inference{}{P^{*}\downarrow}\quad
\inference{P\step{a}P'}{P^{*}\step{a}P'\cdot P^{*}}\\
&\inference{P_1\step{a}P_1'}{\nest{P_1}{P_2}\step{a}P_1'\cdot(\nest{P_1}{P_2})\cdot P_1}\quad
\inference{P_2\step{a}P_2'}{\nest{P_1}{P_2}\step{a}P_2'}\quad
\inference{P_2\downarrow}{\nest{P_1}{P_2}\downarrow}
\enskip.
\end{eqnarray*}
\end{minipage}
}\end{center}
\caption{The operational semantics of nesting and iteration}~\label{fig:semantics-iteration-nesting}
\end{figure}

\FY{Next,} we discuss the phenomenon of forgetfulness. \FY{Bergstra, Bethke and Ponse introduce a process calculus with iteration and nesting~\cite{Bergstra1994,bergstra2001non}} in which a binary nesting operator $\nest{}{}$ and a Kleene star operator ${}^{*}$ are added. In this paper, we add these two operators to \TCP{} \FY{(Strictly speaking, we use an unary variant Kleene star operator)}.
We give the operational semantics of these two operators in Figure~\ref{fig:semantics-iteration-nesting}.

\FY{To get some intuition for the operational interpretation of these operators, note that the processes $P^{*}$ and $\nest{P_1}{P_2}$ respectively satisfy the following equations modulo strong bisimilarity:}
\begin{equation*}
P^{*}= P\cdot P^{*}+\one\quad
\nest{P_1}{P_2}= P_1\cdot (\nest{P_1}{P_2})\cdot P_1+P_2
\enskip
\end{equation*}

Bergstra et al. show how one can specify a half counter using iteration and nesting, which then allows them to
conclude that the behaviour of a Turing machine can be simulated in the calculus with iteration and nesting \FY{(not including recursion)}~\cite{Bergstra1994,bergstra2001non}.

\begin{figure}
\centering
\begin{tikzpicture}[->,>=stealth',shorten >=1pt,auto,node distance=2cm,semithick]
   \tikzstyle{every state}=[rectangle,rounded corners,draw=black, top color=white, bottom color=yellow!50,very thick, inner sep=0.2cm, minimum size=0.7cm, text centered]
    \node[state,initial] (A)                    {$CC_0$};
  \node[state]         (B) [right of=A] {$CC_1$};
  \node[state]         (C) [right of=B] {$CC_2$};
  \node[state]         (D) [right of=C] {$CC_{n-1}$};
  \node[state]         (E) [right of=D] {$CC_{n}$};
  \node[state]         (F) [below of=E]       {$BB_{n}$};
  \node[state]         (G) [left of=F]       {$BB_{n-1}$};
\node[state]         (H) [left of=G]       {$BB_2$};
\node[state]         (I) [left of=H]       {$BB_1$};
\node[state]         (J) [left of=I]       {$BB_0$};
\node[] (K)[right of=E]{};
\node[] (L)[right of=F]{};
  \path (A) edge              node {$a$} (B)
            edge              node {$b$} (J)
        (B) edge            node {$a$} (C)
            edge              node {$b$} (I)
        (C) edge[dashed]              node {} (D)
            edge              node {$b$} (H)
        (D) edge              node {$a$} (E)
            edge              node {$b$} (G)
        (E) edge              node {$b$} (F)
            edge[dashed]      node {} (K)
        (F) edge              node {$a$} (G)
        (G) edge[dashed]                node{} (H)
        (H) edge      node {$a$} (I)
        (I) edge      node {$a$} (J)
        (L) edge[dashed]      node {} (F)
        (J) edge [bend left]            node{$c$} (A);
\end{tikzpicture}
\caption{The transition system of a half counter}~\label{fig:halfcounter}
\end{figure}
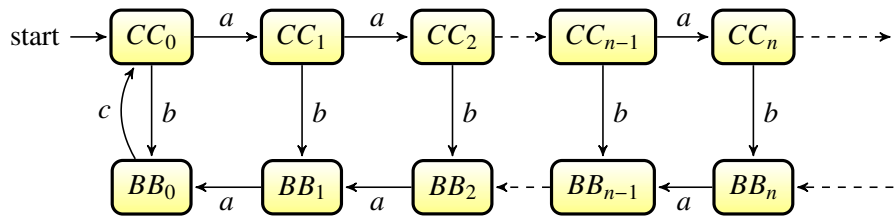

The half counter is specified as follows:
\begin{eqnarray*}
CC_n &=& \mathit{a}.CC_{n+1}+\mathit{b}.BB_{n}\,(n\in\mathbb{N})\\
BB_n &=& \mathit{a}.BB_{n-1}\,(n\geq 1)\\
BB_0 &=& \mathit{c}.CC_0
\enskip.
\end{eqnarray*}

The behaviour of a half counter is illustrated in Figure~\ref{fig:halfcounter}. The initial state is $CC_0$. From $CC_0$ an arbitrary number of $a$ transitions \FY{is possible}. \FY{After} a $b$-labelled transition, \FY{the process performs} the same number of $a$-labelled transitions \FY{as before the $b$-labelled transition,} to the state $BB_0$. In state $BB_0$, a zero testing transition\FY{,} labelled by $c$ \FY{is enabled, leading} back to the state $CC_0$.

An implementation in \FY{a calculus} with iteration and nesting is provided in~\cite{bergstra2001non} as follows:
\begin{equation*}
HCC=((\nest{a}{b})\cdot c)^{*}
\enskip.
\end{equation*}

It is straightforward to establish that $((\nest{a}{b})\cdot a^{n}\cdot c) \cdot HCC$ is equivalent to $CC_n$ for all $n\geq 1$ modulo strong bisimilarity, and $(a^{n}\cdot c)\cdot HCC$ is equivalent to $BB_n$ for all $n\in\mathbb{N}$ modulo strong bisimilarity.

In a context with intermediate termination, one may wonder if it is possible to generalize their result. It is,
however, not clear how to specify an always terminating half counter. At least, a naive generalisation of the
specification of Bergstra et al. does not do the job. The culprit is forgetfulness.
We define a half counter that terminates in every state as follows:
\begin{eqnarray*}
C_n &=& \mathit{a}.C_{n+1}+\mathit{b}.B_{n}+\one\quad(n\in\mathbb{N})\\
B_n &=& \mathit{a}.B_{n-1}+\one\quad(n\geq 1)\\
B_0 &=& \mathit{c}.C_0+\one
\enskip.
\end{eqnarray*}
\FY{Now consider the process $HC$ defined by:}
\begin{equation*}
HC=(\nest{\mathit{(a+\one)}}{\mathit{(b+\one)}}\cdot (c+\one))^{*}
\enskip.
\end{equation*}
Note that due to transparency, $((a+\one)^{n}\cdot (c+\one))\cdot HC$ is \FY{not} equivalent to $B_n$ modulo any \FY{reasonable} behavioural equivalence for $n>1$ since $B_n$ only has an $a$-labelled transition to $B_{n-1}$ whereas the other process has at least $n+1$ transitions leading to $HC,\,(c+\one)\cdot HC,\,(a+\one)\cdot (c+\one)\cdot HC,\,\ldots,\,(a+\one)^{n-1}\cdot (c+\one)\cdot HC$, respectively. This process may choose to ``forget'' the transparent process expressions that have been stacked using the sequential composition operator. We conjecture that, due to
forgetfulness, the always terminating half counter cannot be specified in \TCPN.

In Section~\ref{sec:Termination}, we show that with the revised semantics, it is possible to specify an always terminating half counter and we shall \FY{prove that \TCP{} extended with $*$ and $\sharp$ (but without recursion) is reactively Turing powerful.} 
\section{A Revised Semantics of the Sequential Composition Operator}\label{sec:sequential}
Inspired \FY{by} the work in~\cite{AH1992} and~\cite{Bloom1994}, we \FY{revise the operational semantics for sequential composition and propose a calculus \TCPS}. Its syntax is obtained by replacing the sequential composition operator $\cdot$ by $\bullet$ in the syntax of \TCP{}.
Note that we also use the abbreviation of $P^{n}$ as we did for the standard version of the sequential composition operator.

\FY{The operational rules for $\bullet$ are givem in Figure~\ref{fig:revised-semantics-sequential}.}
\begin{figure}[h]
\begin{center}\fbox{
\begin{minipage}[t]{.8\textwidth}
\begin{eqnarray*}
& \inference{P_1\downarrow\quad P_2\downarrow}{P_1\bullet P_2\downarrow}\quad
\inference{P_1\step{a}P_1'}{P_1\bullet P_2\step{a}P_1'\bullet P_2}\quad \inference{P_1\downarrow\quad P_2\step{a}P_2'\quad P_1\not{\step{}}}{P_1\bullet P_2\step{a}P_2'}
\enskip.
\end{eqnarray*}
\end{minipage}
}\end{center}
\caption{The revised semantics of sequential composition}\label{fig:revised-semantics-sequential}
\end{figure}
\FY{Note that the third rule has a negative premise $P_1\not{\step}$. Intuitively, this rule is only applicable if there does not exist a closed term $P_1'$ and an action $a\in \Atau$ such that the transition $P_1\step{a}P_1'$ is derivable. For a sound formalisation of this intuition, using the notions of irredundant and well-supported proof, see~\cite{vG2004}.} As a consequence, the branching degree of a context-free process is bounded and sequential compositions may have the option to terminate, without being forgetful.

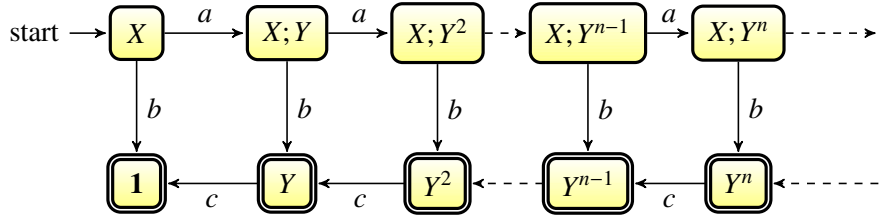
\begin{figure}
\centering
\begin{tikzpicture}[->,>=stealth',shorten >=1pt,auto,node distance=2cm,semithick]
   \tikzstyle{every state}=[rectangle,rounded corners,draw=black, top color=white, bottom color=yellow!50,very thick, inner sep=0.2cm, minimum size=0.7cm, text centered]
       \node[state,initial] (A)                    {$X$};
  \node[state]         (B) [right of=A] {$X\bullet Y$};
  \node[state]         (C) [right of=B] {$X\bullet Y^2$};
  \node[state]         (D) [right of=C] {$X\bullet Y^{n-1}$};
  \node[state]         (E) [right of=D] {$X\bullet Y^{n}$};
  \node[state,accepting]         (F) [below of=E]       {$Y^{n}$};
  \node[state,accepting]         (G) [left of=F]       {$Y^{n-1}$};
\node[state,accepting]         (H) [left of=G]       {$Y^2$};
\node[state,accepting]         (I) [left of=H]       {$Y$};
\node[state,accepting]         (J) [left of=I]       {$\one$};
\node[] (K)[right of=E]{};
\node[] (L)[right of=F]{};
  \path (A) edge              node {$a$} (B)
            edge              node {$b$} (J)
        (B) edge            node {$a$} (C)
            edge              node {$b$} (I)
        (C) edge[dashed]              node {} (D)
            edge              node {$b$} (H)
        (D) edge              node {$a$} (E)
            edge              node {$b$} (G)
        (E) edge              node {$b$} (F)
            edge[dashed]      node {} (K)
        (F) edge              node {$c$} (G)
        (G) edge[dashed]                node{} (H)
        (H) edge      node {$c$} (I)
        (I) edge      node {$c$} (J)
        (L) edge[dashed]      node {} (F);
\end{tikzpicture}
\caption{The transition system in the revised semantics}\label{fig:bounded}
\end{figure}

Let us revisit the first example in Section~\ref{sec:transparency}. We rewrite it with the revised sequential composition operator:
\begin{equation*}
X=a.X\bullet Y+b.\one\quad Y=c.\one+\one
\enskip.
\end{equation*}
Its transition system is illustrated in Figure~\ref{fig:bounded}. Every state in the transition system now has a bounded branching degree. For instance, a transition from $Y^{5}$ to $Y^{2}$ is abandoned because $Y$ has a transition and only the transition from the first $Y$ in the sequential composition is allowed.

Congruence is an important property to fit a behavioural equivalence into an axiomatic framework. We have that in the revised semantics, $\rbbisimd$ is a congruence. Note that the congruence property can also be inferred from a recent result of Fokkink, van Glabbeek and Luttik~\cite{FvGL2017}.
\begin{theorem}~\label{thm:congruence}
$\rbbisimd$ is a congruence with respect to \TCPS.
\end{theorem}

\FY{As a remark, unlike the divergence-preserving variant of rooted branching bisimilarity, the more standard variant that does not require divergence-preservation ($\rbbisim$) is not a congruence for \TCPS{}. Consider
\begin{equation*}
P_1=\tau.\one\quad P_2=(\tau.\one)^{*}\quad Q=a.\one
\enskip.
\end{equation*}
We have $P_1\rbbisim P_2$ but not $P_1\bullet Q\rbbisim P_2\bullet Q$, for $P_1\bullet Q$ can do a $a$-transition after the $\tau$-transition, whereas $P_2\bullet Q$ can only do $\tau$ transitions.
}

We also define a version of \TCP{} with iteration and nesting (\TCPN{}) in the revised semantics.
By removing \FY{the facility of} recursive specification and \FY{the operations $*$ and $\sharp$}, we get \TCPN{}. \conf{\FY{The operational rules for $*$ and $\sharp$ are obtained by replacing, in the rules in Figure~\ref{fig:semantics-iteration-nesting}, all occurrences of $\cdot$ by $\bullet$.}}
\delete{
We extend the calculus with the following operators:
\begin{equation*}
P^{*},\pushd{P_1}{P_2},\backf{P_1}{P_2},\nest{P_1}{P_2}
\end{equation*}

These operations are defined by
\begin{eqnarray*}
\pushd{x}{y}&=& x(\pushd{x}{y})(\pushd{x}{y})+y\\
\backf{x}{y}&=& x(\backf{x}{y})y+y\\
\nest{x}{y}&=& x(\nest{x}{y})x+y
\end{eqnarray*}
}
\arx{
The operational semantics is defined in Figure~\ref{fig:revised-semantics-iteration-nesting}.
\begin{figure}[h]
\fbox{
\begin{minipage}[t]{1\textwidth}
\begin{eqnarray*}
&\inference{}{P^{*}\downarrow}\quad
\inference{P\step{a}P'}{P^{*}\step{a}P'\bullet P^{*}}\\
\delete{&\inference{P_1\step{a}P_1'}{\pushd{P_1}{P_2}\step{a}P_1'\bullet\pushd{P_1}{P_2}\bullet\pushd{P_1}{P_2}}\quad
\inference{P_2\step{a}P_2'}{\pushd{P_1}{P_2}\step{a}P_2'}\quad
\inference{P_2\downarrow}{\pushd{P_1}{P_2}\downarrow}\\
&\inference{P_1\step{a}P_1'}{\backf{P_1}{P_2}\step{a}P_1'\bullet\backf{P_1}{P_2}\bullet P_2}\quad
\inference{P_2\step{a}P_2'}{\backf{P_1}{P_2}\step{a}P_2'}\quad
\inference{P_2\downarrow}{\backf{P_1}{P_2}\downarrow}\\}
&\inference{P_1\step{a}P_1'}{\nest{P_1}{P_2}\step{a}P_1'\bullet(\nest{P_1}{P_2})\bullet P_1}\quad
\inference{P_2\step{a}P_2'}{\nest{P_1}{P_2}\step{a}P_2'}\quad
\inference{P_2\downarrow}{\nest{P_1}{P_2}\downarrow}
\end{eqnarray*}
\end{minipage}
}
\caption{The revised semantics of iteration and nesting}~\label{fig:revised-semantics-iteration-nesting}
\end{figure}
} 
\section{Context-free Processes and Pushdown Processes}\label{sec:cfg}
The relationship between context-free processes and pushdown processes has been studied in the literature~\cite{BCvT2008}.
We consider the process calculus \emph{Theory of Sequential Processes} (\SAS). We \FY{define} context-free processes as follows:

\begin{definition}~\label{def:cfp}
A \emph{context-free process} is the \FY{strong} bisimulation equivalence class of the transition system generated by a finite guarded recursive specification over \SAS.
\end{definition}

Note that there is a method to rewrite every context-free process into Greibach normal form~\cite{BBK93}, which is also valid in the revised semantics. In this paper, we only consider context-free processes in Greibach normal form \FY{,i.e., defined by guarded recursive specifications of the form}
\begin{equation*}
X=\sum_{i\in I_X}\alpha_i.\xi_i(+\one)
\enskip.
\end{equation*}

In this form, every right-hand side of every equation consists of a number of summands, indexed by a finite set $I_X$ (the empty sum \FY{denotes} $\nil$), each of which is $\one$, or of the form $\alpha_i.\xi_i$, where $\xi_i$ is the sequential composition of names (the empty sequence \FY{denotes} $\one$). \delete{We define $I$ as the multiset resulting of the union of all index sets. For a recursive specification in Greibach normal form, every state of the transition system is given by a sequence of names. Note that we can take the index sets associated with the names to be disjoint, so that we can define a function $V : I\rightarrow \V$ that gives, for any index that occurs somewhere in the specification, the name of the equation in which it occurs.}

We shall show that every context-free process is equivalent to a pushdown process modulo strong bisimilarity. The notion of pushdown \FY{automaton} is defined as follows:

\begin{definition}~\label{def:pda}
A pushdown automaton (\PDA) is a $7$-tuple $(\Sta,\Sigma,\D,\step{},\uparrow,Z,\downarrow)$, where
\begin{enumerate}
\item $\Sta$ is a finite set of \emph{states},
\item $\Sigma$ is a finite set of \emph{input symbols},
\item $\D$ is a finite set of \emph{stack symbols},
\item ${\step{}}\subseteq \Sta\times\D\times\Sigma\times\D^{*}\times\Sta$ is a finite \emph{transition relation}, (we write $s\step{a[d/\delta]}t$ for $(s,d,a,\delta,t)\in{\step{}}$),
\item ${\uparrow}\in\Sta$ is the \emph{initial state},
\item $Z\in\D$ is the \emph{initial stack symbol}, and
\item ${\downarrow}\subseteq\Sta$ is a set of \emph{accepting states}.
\end{enumerate}
\end{definition}

We use a sequence of stack symbols $\delta\in\D^{*}$ to represent the contents of a stack. We associate with every pushdown automaton a labelled transition system. The bisimulation equivalence classes of transition systems associated with pushdown automata are referred \FY{to} as \emph{pushdown processes}.
\begin{definition}~\label{def:lts-pda}
   Let $\M=(\Sta,\Sigma,\D,\step{},\uparrow,Z,\downarrow)$ be a \PDA. The \emph{transition system} $\T(\M)=(\Sta_{\T},\step{}_{\T},\uparrow_{\T},\downarrow_{\T})$ associated with $\M$ is defined as follows:
\begin{enumerate}
\item its set of states is the set $\Sta_{\T}= \{(s,\delta)\mid s\in\Sta,\delta\in\D^{*}\}$ of all configurations of $\M$,
\item its transition relation ${\step{}_{\T}}\FY{\subseteq} \Sta_{\T}\times \Atau\times \Sta_{\T}$ is \FY{the} relation satisfying, for all $a\in\Sigma$, $d\in\D$, $\delta,\delta'\in\D^{*}$: $(s,d\delta)\step{a}_{\T}(t,\delta'\delta)$ iff $s\step{a[d/\delta']}t$,
\item its initial state is the configuration $\uparrow_{\T}=(\uparrow,Z)$, and
\item its set of \FY{terminating} states is the set $\downarrow_{\T}=\{(s,\delta)\mid s\in\Sta,\,s\downarrow,\,\delta\in\D^{*}\}$.
\end{enumerate}
\end{definition}

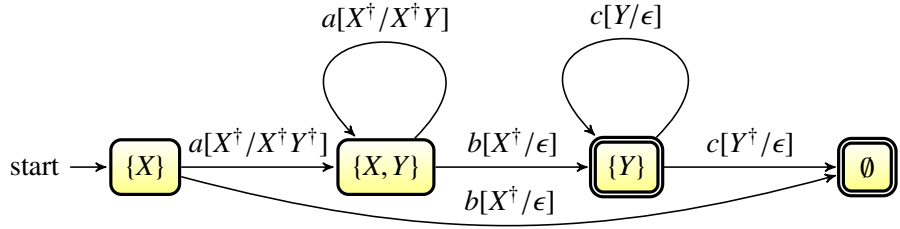
\begin{figure}
\centering
\begin{tikzpicture}[->,>=stealth',shorten >=1pt,auto,node distance=3.2cm,semithick]
   \tikzstyle{every state}=[rectangle,rounded corners,draw=black, top color=white, bottom color=yellow!50,very thick, inner sep=0.2cm, minimum size=0.7cm, text centered]
  \node[state,initial] (A)                    {$\{X\}$};
  \node[state]         (B) [right of=A] {$\{X,Y\}$};
  \node[state,accepting]         (C) [right of=B] {$\{Y\}$};
  \node[state,accepting]         (D) [right of=C] {$\emptyset$};

  \path (A) edge              node {$a[X^{\dagger}/X^{\dagger}Y^{\dagger}]$} (B)
            edge [bend right=15]             node {$b[X^{\dagger}/\epsilon]$} (D)
        (B) edge [loop]           node[above] {$a[X^{\dagger}/X^{\dagger}Y]$} (B)
            edge              node {$b[X^{\dagger}/\epsilon]$} (C)
        (C) edge [loop]            node[above] {$c[Y/\epsilon]$} (C)
            edge              node {$c[Y^{\dagger}/\epsilon]$} (D)
;
\end{tikzpicture}
\caption{A PDA to simulate the process in Figure~\ref{fig:bounded}}\label{fig:example-pda}
\end{figure}

\FY{Recall that a context-free process is defined by a recursive specification in Greibach normal form; all states of the
context-free process are denoted by sequences of names defined in this recursive specification. Note that a sequence of names denotes a
terminating state only if all names have the option to terminate. Hence, to be able to determine whether a configuration of the pushdown automaton should have the option to terminate, we need to know whether all names currently on the stack have the option to terminate. We annotate the states of the pushdown automaton with the subset of names currently on the stack. We shall use the stack to record the sequence of names corresponding to the current state.} The \FY{deepest} occurrence of a name \FY{on the stack} is marked and we shall include special transitions in the automaton \FY{for the treatment of} marked names. If a marked name is removed from the stack, then \FY{, intuitively, it should be removed from the set annotating the state} from the set. On the other hand, if a name not in the set is added to the stack, then we shall mark that name and add that name to the set \FY{annotating the state}. As an example, we introduce a \PDA{}  \delete{\FY{$\M=(\Sta,\Sigma,\D,\step{},\uparrow,Z,\downarrow)$}} \FY{as in Figure~\ref{fig:example-pda}} to simulate the process in Figure~\ref{fig:bounded} modulo $\bisim$. \delete{as follows: $\Sta=\{{\emptyset},{\{X\}},{\{Y\}},{\{X,Y\}}\}$, $\Sigma=\{a,b,c\}$, $\D=\{X,Y,X^{\dagger},Y^{\dagger}\}$, $Z=X^{\dagger}$, $\uparrow={\{X\}}$, $\downarrow=\{{\emptyset},{\{Y\}}\}$, and $\step{}=\{({\{X\}},X^{\dagger},a,X^{\dagger}Y^{\dagger},{\{X,Y\}}), ({\{X\}},X^{\dagger},b,\epsilon,\emptyset),
({\{X,Y\}},X^{\dagger},a,X^{\dagger}Y,{\{X,Y\}}),\\
({\{X,Y\}},X^{\dagger},b,\epsilon,{\{Y\}}),
({\{Y\}},Y,c,\epsilon,{\{Y\}}),
({\{Y\}},Y^{\dagger},c,\epsilon,{\emptyset}),\}$.}
\delete{
\begin{enumerate}
\item $\Sta=\{{\emptyset},{\{X\}},{\{Y\}},{\{X,Y\}}\}$;
\item $\Sigma=\{a,b,c\}$;
\item $\D=\{X,Y,X^{\dagger},Y^{\dagger}\}$
\item $\step{}=\{({\{X\}},X^{\dagger},a,X^{\dagger}Y^{\dagger},{\{X,Y\}}),({\{X\}},X^{\dagger},b,\epsilon,\emptyset),
({\{X,Y\}},X^{\dagger},a,X^{\dagger}Y,{\{X,Y\}}),
({\{X,Y\}},X^{\dagger},b,\epsilon,{\{Y\}}),\\
({\{Y\}},Y,c,\epsilon,{\{Y\}}),
({\{Y\}},Y^{\dagger},c,\epsilon,{\emptyset}),\}$;
\item $\uparrow={\{X\}}$;
\item $Z=X^{\dagger}$; and
\item $\downarrow=\{{\emptyset},{\{Y\}}\}$.
\end{enumerate}
}

To obtain a general result, we consider a context-free process defined by a set of names $\V=\{X_0,X_1,\ldots,X_m\}$ with $X_0$ as the initial state, where
\begin{equation*}
X_j=\sum_{i\in I_{X_j}}\alpha_{ij}.\xi_{ij}(+\one)
\enskip.
\end{equation*}

We introduce the following \FY{auxiliary} functions:
\begin{enumerate}
\item $\mathit{length}:\V^{*}\rightarrow\mathbb{N}$, $\length{\xi}$ is the length of $\xi$;
\item $\mathit{get}:\V^{*}\times\mathbb{N}\rightarrow \V$, $\get{\xi}{i}$ is the $i$-th name of $\xi$;
\item $\mathit{suffset}:\V^{*}\times\mathbb{N}\rightarrow 2^{\lvert \V\rvert}$, $\suffset{\xi}{i}=\{\get{\xi}{j}\mid j=i+1,\ldots\length{\xi}\}$ computes the set that contains all the names in the suffix which starts from the $i$-th name of $\xi$.
\end{enumerate}
We define a \PDA{} $\M=(\Sta,\Sigma,\D,\step{},\uparrow,Z,\downarrow)$ to simulate the transition system associated with $X_0$ as follows:
\FY{
$\Sta=\{D\mid D\subseteq \V\}$; $\Sigma=\Atau$; $\D=\V\cup\{X^{\dagger}\mid X\in\V\}$; $\uparrow={\{X_0\}}$; $Z=X^{\dagger}_{0}$;
$\downarrow=\{D\mid \mbox{if for all}\,X\in D, X\downarrow\}$;} and the \FY{transition relation} $\step{}$ is defined as follows:
\begin{eqnarray*}
\step{}&=&\{(D,X^{\dagger}_{j},\alpha_{ij},\delta(D,X^{\dagger}_{j},\xi_{ij}),{\FY{\mer}(D,X^{\dagger}_{j},\xi_{ij})})\mid i\in I_{X_j},\,j=1,\ldots,n,\,D\FY{\subseteq}\V\}\\
&\cup&\{(D,X_j,\alpha_{ij},\delta(D,X_j,\xi_{ij}),{\FY{\mer}(D,X_j,\xi_{ij})})\mid i\in I_{X_j},\,j=1,\ldots,n,\,D\FY{\subseteq}\V\}
\enskip,
\end{eqnarray*}
where $\delta(D,X^{\dagger}_{j},\xi_{ij})$ is \FY{the} string of length $\length{\xi_{ij}}$ defined as follows: for $k=1,\ldots, \length{\xi_{ij}}$, we let $X_k=\get{\xi_{ij}}{k}$,
\begin{enumerate}
 \item if $X_k\notin (D/\{X_j\})\cup\suffset{\xi_{ij}}{k}$, then the $k$-th symbol of $\delta(D,X^{\dagger}_{j},\xi_{ij})$ is $X^{\dagger}_{k}$,
\item otherwise, the $k$-th symbol of $\delta(D,X^{\dagger}_{j},\xi_{ij})$ is $X_k$,
\end{enumerate}
$\delta(D,X_j,\xi_{ij})$ is a string of length $\length{\xi_{ij}}$ defined as follows: for $k=1,\ldots,\length{\xi_{ij}}$, we let $X_k=\get{\xi_{ij}}{k}$,
\begin{enumerate}
 \item if $X_{k}\notin D\cup\suffset{\xi_{ij}}{k}$, then the $k$-th symbol of $\delta(D,X_j,\xi_{ij})$ is $X^{\dagger}_k$,
\item otherwise, the $k$-th symbol of $\delta(D,X_j,\xi_{ij})$ is $X_k$, and
\end{enumerate}
we also define $\FY{\mer}(D,X^{\dagger}_j,\xi_{ij})=(D/\{X_j\})\cup\suffset{\xi_{ij}}{0}$ and $\FY{\mer}(D,X_j,\xi_{ij})=D\cup\suffset{\xi_{ij}}{0}$;

We \FY{have} the following result:
\begin{lemma}~\label{lemma:cfp-pda}
$\T(X_0)\bisim \T(\M)$.
\end{lemma}

We have the following theorem.
\begin{theorem}~\label{thm:cfp-pda}
\FY{For every name $X$ defined in a guarded recursive specification in Greibach normal form there exists a \PDA{} $\M$, such that $\T(X)\bisim \T(\M)$.}
\end{theorem}

Note that the converse of this theorem does not hold in general, a counterexample was established by F. Moller in~\cite{Moller1996}, and we \FY{conjecture} that it is also valid modulo $\bbisim$ in the revised semantics.

\section{Executability in the Context of Termination}\label{sec:Termination}
\delete{In this section, we shall discuss the theory of executability in the context of termination. We shall prove that \TCPN{} is reactively Turing powerful in the context of termination.
}
The notion of reactive Turing machine (RTM)~\cite{BLT2013} was introduced as an extension of Turing machines to define which behaviour is executable by a computing system.
 The definition of RTM is parameterised with the set $\Atau$, which we now assume to be finite, and with another finite set $\D$ of \emph{data symbols}. We extend $\D$ with a special symbol $\Box\notin\D$ to denote a blank tape cell, and denote the set $\D\cup\{\Box\}$ of \emph{tape symbols} by $\Dbox$.
\begin{definition}
[Reactive Turing Machine]\label{def:rtm}
A \emph{reactive Turing machine} (RTM) is a quadruple $(\Sta,\step{},\uparrow,\downarrow)$, where
\begin{enumerate}
    \item $\Sta$ is a finite set of \emph{states},
    \item ${\step{}}\subseteq \Sta\times\Dbox\times\Atau\times\Dbox\times\{L,R\}\times\Sta$ is a finite collection of $(\Dbox\times\Atau\times\Dbox\times\{L,R\})$-labelled \emph{transitions} (we write $s\step{a[d/e]M}t$ for $(s,d,a,e,M,t)\in{\step{}}$),
    \item ${\uparrow}\in\Sta$ is a distinguished \emph{initial state}, and
    \item ${\downarrow}\subseteq\Sta$ is a finite set of \emph{final states}.
\end{enumerate}
\end{definition}

Intuitively, the meaning of  a transition $s\step{a[d/e]M}t$ is that whenever the RTM is in state $s$, and $d$ is the symbol currently read by the tape head, then it may execute the action $a$, write symbol $e$ on the tape (replacing $d$), move the read/write head one position to the left or the right on the tape (depending on whether $M=L$ or $M=R$), and then end up in state $t$.

To formalise the intuitive understanding of the operational behaviour of RTMs, we associate with every RTM $\M$ an $\Atau$-labelled transition system  $\T(\M)$. The states of $\T(\M)$ are the configurations of $\M$, which consist of a state from $\Sta$, its tape contents, and the position of the read/write head.
We denote by $\check{\Dbox}=\{\check{d}\mid d\in\Dbox\}$ the set of \emph{marked} symbols; a \emph{tape instance} is a sequence $\delta\in(\Dbox\cup\check{\Dbox})^{*}$ such that $\delta$ contains exactly one element of the set of marked symbols $\check{\Dbox}$, indicating the position of the read/write head.
We adopt a convention to concisely denote an update of the placement of the tape head marker. Let $\delta$ be an element of $\Dbox^{*}$. Then by $\tphdL{\delta}$ we denote the element of $(\Dbox\cup\check{\Dbox})^{*}$ obtained by placing the tape head marker on the right-most symbol of $\delta$ (if that exists; otherwise $\tphdL{\delta}$ denotes $\tphd{\Box}$).
Similarly, $\tphdR{\delta}$ is obtained by placing the tape head marker on the left-most symbol of $\delta$ (if that exists; otherwise $\tphdR{\delta}$ denotes $\tphd{\Box}$).
\begin{definition}\label{def:lts-tm}
Let $\M=(\Sta,\step{},\uparrow,\downarrow)$ be an RTM. The \emph{transition system} $\T(\M)$ \emph{associated with} $\M$ is defined as follows:
\begin{enumerate}
\item its set of states is the set $\Conf[\M]=\{(s,\delta)\mid s\in\Sta,\ \text{$\delta$ a tape instance}\}$ of all configurations of $\M$;
    \item its transition relation ${\step{}}\subseteq{\Conf[\M]\times\Atau\times\Conf[\M]}$ is the relation satisfying, for all $a\in\Atau,\,d,e\in\Dbox$ and $\delta_L,\delta_R\in\Dbox^{*}$:
      $(s,\delta_L\check{d}\delta_R)\step{a}(t,\tphdL{\delta_L}e\delta_R)$ iff $s\step{a[d/e]L}t$, and
        $(s,\delta_L\check{d}\delta_R)\step{a}(t,\delta_L e{}\tphdR{\delta_R})$ iff $s\step{a[d/e]R}t$;

    \item its initial state is the configuration $(\uparrow,\check{\Box})$; and
    \item its set of final states is the set $\{(s,\delta)\mid \text{$\delta$ a tape instance},\, s\downarrow\}$.
\end{enumerate}
\end{definition}

Turing introduced his machines to define the notion of \emph{effectively computable function} in~\cite{Turing1936}. By analogy, the notion of RTM can be used to define a notion of \emph{effectively executable behaviour}.

\begin{definition}
[Executability]\label{def:exe}
A transition system is \emph{executable} if it is the transition system associated with some RTM.
\end{definition}

Executability can be used to characterise the absolute expressiveness of process calculi in two \FY{ways}. On the one hand, if every transition system associated with a process expression specified in a process calculus is executable modulo some behavioural equivalence, then we say that the process calculus is \emph{executable} modulo that behavioural equivalence. On the other hand, if every executable transition system is behaviourally equivalent to some transition system associated with a process expression specified in a process calculus modulo some behavioural equivalence, then we say that the process calculus is \emph{reactively Turing powerful} modulo that behavioural equivalence.

Our aim in this section is to prove that all executable processes can be specified, up to divergence-preserving branching bisimilarity in \TCPN. \TCPN{} is obtained from \TCP{} by removing recursive \FY{definitions} and adding the iteration and nesting \FY{operators}.

\FY{To see} that \TCPN{} \FY{is} executable \FY{modulo branching bisimilarity, it suffices to} observe that their transition systems are \FY{effective}. Thus we can apply the result from~\cite{BLT2013} and \FY{conclude} that they are executable modulo $\bbisim$.

Now we show that \TCPN{} is reactively Turing powerful modulo $\bbisimd$.

 We first introduce the notion of bisimulation up to $\bbisim$, which is a useful tool to establish the proofs in this section. Note that we adopt a non-symmetric bisimulation up to relation.

\begin{definition}\label{def:up-to}
Let $T=(\Sta,\step{},\uparrow,\downarrow)$ a transition system. A relation $\R\subseteq\Sta\times\Sta$ is a bisimulation up to $\bbisim$ if, whenever $s_1\R s_2$, then for all $a\in \Atau$:
\begin{enumerate}
    \item if $s_1\step{}^{*}s_1''\step{a}s_1'$, with $s_1\bbisim s_1''$ and ${a\neq\tau}\vee{s_1''\not\bbisim s_1'}$, then there exists $s_2'$ such that $s_2\step{a}s_2'$, $s_1''\mathrel{\bbisim\mathrel{\circ}\mathrel{\R}}s_2$ and $s_1'\mathrel{\bbisim \mathrel{\circ} \mathrel{\R}} s_2'$;
    \item if $s_2\step{a}s_2'$, then there exist $s_1',s_1''$ such that $s_1\step{}^{*}s_1''\step{a}s_1'$, $s_1''\bbisim s_1$ and $s_1'\mathrel{\bbisim \mathrel{\circ} \mathrel{\R}} s_2'$;
    \item if $s_1\downarrow$, then there exists $s_2'$ such that $s_2\step{}^{*} s_2'$\FY{, $s_2'\downarrow$ and $s_1\R s_2'$}; and
    \item if $s_2\downarrow$, then there exists $s_1'$ such that $s_1\step{}^{*} s_1'$\FY{, $s_1'\downarrow$ and $s_1'\R s_2$}.
\end{enumerate}
\end{definition}

\begin{lemma}\label{lemma:up-to}
If $\R$ is a bisimulation up to $\bbisim$, then $\R \subseteq {\bbisim}$.
\end{lemma}

Next we show that \TCPN{} is reactively Turing powerful by writing a specification of the transition system associated with a reactive Turing machine in \TCPN{} modulo $\bbisimd$. The proof proceeds in five steps:
\begin{enumerate}
\item We first specify an always terminating half counter.
\item Then we show that every regular process can be specified in \TCPN.
\item Next we use two half counters and a regular process to encode a terminating stack.
\item With two stacks and a regular process we can specify a tape.
\item Finally we use a tape and a regular control process to specify an RTM.
\end{enumerate}

We first recall the infinite specification in \SAS{} of a terminating half counter from Section~\ref{sec:transparency}.
\arx{\begin{eqnarray*}
C_n &=& \mathit{a}.C_{n+1}+\mathit{b}.B_{n}+\one\,(n\in\mathbb{N})\\
B_n &=& \mathit{a}.B_{n-1}+\one\,(n\geq 1)\\
B_0 &=& \mathit{c}.C_0+\one
\enskip.
\end{eqnarray*}
}
We provide a specification of a counter in \TCPN{} as follows:
\begin{equation*}
HC=(\nest{\mathit{(a+\one)}}{\mathit{(b+\one)}}\bullet (c+\one))^{*}
\end{equation*}

We have the following lemma:

\begin{lemma}~\label{lemma:tcpn-halfcounter}
$C_0\bbisimd HC$.
\end{lemma}

Next we show that every regular process can be specified in \TCPN{} modulo $\bbisimd$.
A regular process is given by $P_i=\sum_{j=1}^{n} \alpha_{ij}\bullet P_j +\beta_i\,(i=1,\ldots,n)$ where $\alpha_{ij}$ and $\beta_{i}$ are finite sums of actions from $\Atau$ and possibly with a $\one$-summand. We have the following lemma.

\begin{lemma}~\label{lemma:tcpn-rular}
Every regular process can be specified in \TCPN{} modulo $\bbisimd$.
\end{lemma}

Now we show that a stack can be specified by a regular process and two half counters. We first give an infinite specification in \SAS{} of a stack as follows:
\begin{eqnarray*}
S_{\epsilon}&=&\Sigma_{d\in\Dbox} \mathit{push}?d.S_{d}+\mathit{pop}!\Box.S_{\epsilon}+\one\\
S_{d\delta}&=& \mathit{pop}!d.S_{\delta}+\Sigma_{e\in\Dbox} \mathit{push}?e.S_{ed\delta}+\one
\enskip.
\end{eqnarray*}

Note that $\Dbox$ is a finite set of symbols. We suppose that $\Dbox$ contains $N$ symbols (including $\Box$).
We use $\epsilon$ to denote the empty sequence.
We inductively define an encoding from a sequence of symbols to a natural number $\encode{\_}:{\Dbox}^{*}\FY{\rightarrow}\mathbb{N}$ as follows:
\begin{equation*}
\encode{\epsilon}=0\quad
\encode{d_k}=k\quad(k=1,2,\ldots,N)\quad
\encode{d_k\sigma}=k+N\times\encode{\sigma}
\enskip.
\end{equation*}
Hence we are able to encode the contents of a stack in terms of natural numbers recorded by half counters. We define a stack in \TCPN{} as follows:
\begin{eqnarray*}
S&=&[X_{\emptyset}\parallel P_1\parallel P_2]_{\{a_1,a_2,b_1,b_2,c_1,c_2\}}\\
P_j&=&(\nest{(a_j!a+\one)}{(b_j!b+\one)}\bullet(c_j!c+\one))^{*}\,(j=1,2)\\
X_{\emptyset}&=&(\Sigma_{j=1}^{N}((\mathit{push}?d_j+\one)\bullet(a_1?a+\one)^{j}\bullet(b_1+\one)\bullet X_j)+\mathit{pop!\Box})^{*}\\
X_{k}&=&\Sigma_{j=1}^{N}((\mathit{push}?d_j+\one)\bullet \mathit{Push_j})+(\mathit{pop}!d_k+\one)\bullet \mathit{Pop_k}\quad(k=1,2,\ldots,N)\\
\mathit{Push_k}&=&\mathit{Shift1to2}\bullet(a_1?a+\one)^k\bullet\mathit{NShift2to1}\bullet X_{k}\quad(k=1,2,\ldots,N)\\
\mathit{Pop_k}&=&(a_1?a+\one)^k\bullet\mathit{1/NShift1to2}\bullet\mathit{Test_{\emptyset}}\\
\mathit{Shift1to2}&=&((a_1?a+\one)\bullet(a_2?a+\one))^{*}\bullet(c_1?c+\one)\bullet(b_2?b+\one)\\
\mathit{NShift2to1}&=&((a_2?a+\one)\bullet(a_1?a+\one)^{N})^{*}\bullet(c_2?c+\one)\bullet(b_1?b+\one)\\
\mathit{1/NShift1to2}&=&((a_1?a+\one)^{N}\bullet(a_2?a+\one))^{*}\bullet(c_1?c+\one)\bullet(b_2?b+\one)\\
\mathit{Test_{\emptyset}}&=&(a_2?a+\one)\bullet(a_1?a+\one)\bullet\mathit{Test_1}+(c_2?c+\one)\bullet X_{\emptyset}\\
\mathit{Test_{1}}&=&(a_2?a+\one)\bullet(a_1?a+\one)\bullet\mathit{Test_2}+(c_2?c+\one)\bullet X_{1}\\
\mathit{Test_{2}}&=&(a_2?a+\one)\bullet(a_1?a+\one)\bullet\mathit{Test_3}+(c_2?c+\one)\bullet X_{2}\\
&\cdots&\\
\mathit{Test_{N}}&=&(a_2?a+\one)\bullet(a_1?a+\one)\bullet\mathit{Test_1}+(c_2?c+\one)\bullet X_{N}
\enskip.
\end{eqnarray*}

We have the following result.
\begin{lemma}~\label{lemma:tcpn-stack}
$S_{\epsilon}\bbisimd S$.
\end{lemma}

Next we proceed to define the tape by means of two stacks. We consider the following infinite specification in \SAS{} of a tape:
\begin{equation*}
T_{\delta_L\tphd{d}\delta_R}=r!d.T_{\delta_L\tphd{d}\delta_R}+\Sigma_{e\in \Dbox}w?e.T_{\delta_L\tphd{e}\delta_R}+L?m. T_{\tphdL{\delta_L}d\delta_R}+R?m.T_{\delta_Ld\tphdR{\delta_R}}+\one
\enskip.
\end{equation*}

We define the tape process in \TCPN{} as follows:
\begin{eqnarray*}
T&=&[T_{\Box}\parallel S_1\parallel S_2]_{\{\mathit{push_1,pop_1,push_2,pop_2}\}}\\
T_{d}&=&r!d.T_{d}+\Sigma_{e\in\Dbox}w?e.T_{e}+L?m.\mathit{Left_d}+R?m.\mathit{Right_d}+\one\quad (d\in\Dbox)\\
\mathit{Left_d}&=&\Sigma_{e\in\Dbox}((\mathit{pop_1?e}+\one)\bullet(\mathit{push_2!d}+\one)\bullet T_{e})\\
\mathit{Right_d}&=&\Sigma_{e\in\Dbox}((\mathit{pop_2?e}+\one)\bullet(\mathit{push_1!d}+\one)\bullet T_{e})
\enskip,
\end{eqnarray*}
where $S_1$ and $S_2$ are two stacks obtained by renaming $\mathit{push}$ and $\mathit{pop}$ in $S$ to $\mathit{push_1,pop_1,push_2}$ and $\mathit{pop_2}$, respectively.
We establish the following result.
\begin{lemma}~\label{lemma:tcpn-tape}
$T_{\tphd{\Box}}\bbisimd T$.
\end{lemma}

Finally, we construct a finite control process for an RTM $\M=(\Sta_{\M},\step{}_{\M},\uparrow_{\M},\downarrow_{\M})$ as follows:
\begin{equation*}
C_{s,d}=\Sigma_{(s,d,a,e,M,t)\in\step{}_{\M}}(a.w!e.M!m.\Sigma_{f\in\Dbox}r?f.C_{t,f})[+\one]_{s\downarrow_{\M}}\,(s\in\Sta_{\M},d\in\Dbox)
\enskip.
\end{equation*}

We prove the following lemma.
\begin{lemma}~\label{lemma:tcpn-control}
$\T(\M)\bbisimd [C_{\uparrow_{\M},\Box}\parallel T]_{\{r,w,L,R\}}$.
\end{lemma}

We have the following theorem.

\begin{theorem}~\label{thm:tcpn}
\TCPN{} is reactively Turing powerful modulo $\bbisimd$.
\end{theorem} 
\section{Conclusion}\label{sec:conclusion}

The results established in this paper show that a revision of the operational semantics of sequential composition leads to a smoother integration of process theory and the classical theory of automata and formal languages. In particular, the correspondence between context-free processes and pushdown processes can be established up to strong bisimilarity, which does not hold with the more standard operational semantics  of sequential composition in a setting with intermediate termination \cite{BBR2010}. Furthermore, the revised operational semantics of sequential composition also seems to work better in combination with the recursive operations of \cite{bergstra2001non}. We conjecture that it is not possible to specify an always terminating counter or stack in a process calculus with iteration and nesting if the original operational semantics of sequential composition is used.

There are also some disadvantages to the revised operational semantics.

First of all, the negative premise in the operational semantics gives well-known formal complications in determining whether some process does, or does not, admit a transition. For instance, consider the following unguarded recursive specification:
\begin{equation*}
X=X\bullet Y+\one\quad
Y=a.\one
\enskip.
\end{equation*}
It is not a priori clear whether an $a$-transition is possible from $X$: if $X$ \emph{only} has the option to terminate, then $X\bullet Y$ can do the $a$-transition from $Y$, but then also $X$ can do the $a$-transition, contradicting the assumption that $X$ \emph{only} has the option to terminate.

\FY{Second, as we have illustrated in Section~\ref{sec:sequential}, rooted branching bisimilarity is not compatible with respect to the new sequential composition operation. The divergence-preserving condition is required for the congruence property.}

Finally, note that $(a+\one)\bullet b$ is not strongly bisimilar to $(a\bullet b)+(\one\bullet b)$, and hence $;$ does not distribute from the right over $+$. It is to be expected that there is no finite sound and ground-complete set of equational axioms for the process calculus \TCPS{} with respect to strong bisimilarity. We leave for future work to further investigate the equational theory of sequential composition.

Another interesting future work is to establish the reactive Turing powerfulness on other process calculi with non-regular iterators based on the revised semantics of the sequential composition operator. For instance, we could consider the pushdown operator ``$\$$'' and the back-and-forth operator ``$\leftrightarrows$'' introduced by Bergstra and Ponse in~\cite{bergstra2001non}. They are given by the following equations:
\begin{equation*}
\pushd{P_1}{P_2}= P_1\bullet (\pushd{P_1}{P_2})\bullet (\pushd{P_1}{P_2})+P_2\quad
\backf{P_1}{P_2}= P_1\bullet (\backf{P_1}{P_2})\bullet P_2+P_2
\enskip.
\end{equation*}
By analogy to the nesting operator, we shall also give them some proper rules of operational semantics, and then use the calculus obtained by the revised semantics to define other versions of terminating counters. In a way, we should be able to establish their reactive Turing powerfulness.

\bibliographystyle{eptcs}
\bibliography{Executability}
\delete{\conf{
\appendix
\input{Appendix}
}}
\end{document}